# Enhanced van der Waals epitaxy via electron transfer-enabled interfacial dative bond formation

Weiyu Xie[1], Toh-Ming Lu[1], Gwo-Ching Wang[1], Ishwara Bhat[2], and Shengbai Zhang[1]*

[1]*Department of Physics, Applied Physics & Astronomy,*

*Rensselaer Polytechnic Institute, Troy, NY 12180, USA*

[2]*Department of Electrical, Computer and Systems Engineering,*

*Rensselaer Polytechnic Institute, Troy, NY 12180, USA*

*Corresponding author, Shengbai Zhang. Phone: (518) 276-6127; Email: zhangs9@rpi.edu

**Abstract**

Enhanced van der Waals (vdW) epitaxy of semiconductors on layered vdW substrate is identified as the formation of dative bonds. For example, despite that $NbSe_2$ is a vdW layered material, first-principles calculations reveal that the bond strength at CdTe-$NbSe_2$ interface is five times as large as that of vdW interaction at CdTe-graphene interface. The unconventional chemistry here is enabled by an effective net electron transfer from Cd dangling-bond states at CdTe surface to metallic non-bonding $NbSe_2$ states, which is a necessary condition to activate the Cd for enhanced binding with Se.

Epitaxial growth of materials with a large lattice and/or crystal symmetry mismatch on substrates via weak van der Waals (vdW) forces [1] has recently attracted much attention [2], especially after the concept of two-dimensional (2D) heterojunctions was proposed by Geim [3]. By definition, vdW epitaxy requires that both in-plane and out-plane orientations are well aligned between the overgrown materials and underlying substrates through vdW binding. To date, there exist ample examples of successful vdW epitaxy of a 2D layered vdW material on another 2D layered vdW substrate [4-7]. It is attempting to apply the same concept to the growth of ordinary semiconductors of a 3D crystal structure on a layered substrate. Since one can mechanically peel off or chemically etch off a thin 2D layer and then place it on another substrate [8], the success of 3D on 2D growth could redefine the scope and applicability of epitaxy for applications not possible before. For example, a number of studies have been undertaken to achieve high quality 3D nanowires (NWs) on 2D layered vdW substrates [9-14].

However, towards the goal of growing 3D materials with a continuous film morphology instead of the discrete NWs, the success is limited. Currently, the already-realized 3D-on-2D vdW epitaxy systems possess a spread of misorientation alignment in contrast to 2D-on-2D, unless with well-controlled growth techniques, e.g., the step-flow growth of single-crystal GaN on a "ragged" graphene [15]. This is because vdW binding is weak and the resulting energy landscape may not exhibit clearly-defined minima as a function of the in-plane orientation angle that is a must for high-quality epitaxy. For example, in the recent vdW epitaxy experiment of CdTe on graphene [16], the measured full-width-at-half-maximum (FWHM) in-plane orientation dispersion is ~14°. In contrast, a significantly smaller FWHM around 5° was found for CdTe-on-$NbSe_2$ [17]. This noticeable difference here is quite surprising because, judged from interlayer binding of the layered structures, both $NbSe_2$ and graphene are typical vdW materials [18].

Theory based on the idea of coincident-site lattice (CSL) matching have been applied to explain the 2D-on-2D vdW epitaxy [7, 19]. The epitaxial relationship is realized here not only by the interfacial interaction, but also by the lattice matching condition, i.e., by a minimization of the lattice strains. In 3D-on-2D vdW epitaxy, however, the same CSL matching assumption no longer holds as lattice-strain energy will increase with the thickness of the 3D material. To overcome such a difficulty, a cluster model has been used to simulate the growth of 3D structure. In general, however, there is no theory that can apply to the vdW epitaxy of a 3D solid on a 2D substrate nor to the understanding of substrate dependence observed experimentally.

In this paper, we develop a first-principles theory for 3D vdW epitaxy on 2D substrate. By a comparative study of CdTe-on-NbSe$_2$ and CdTe-on-Graphene (G), the physics of significantly enhanced structural order in vdW epitaxy is elucidated. We find that when the dangling-bond electrons on the Cd atoms at the contacting surface are transferred to non-bonding NbSe$_2$ states, the formation of directional Cd-Se dative bonds is activated. The interfacial bond strength of CdTe-on-NbSe$_2$ is about five times that of CdTe-on-G, but is significantly weaker than the usual Cd-Se bonds. As such, the enhanced binding can significantly reduce in-plane orientation dispersion, while keeping the highly desirable incommensurate and defect-free properties of the epitaxial films. While NbSe$_2$ is a metallic substrate, we expect the same principle for enhanced vdW epitaxy may also apply to semiconducting substrate by *p*-type doping.

A detailed account of first-principles theories used in the study is given in the Supplemental Material along with the formulation of the interfacial formation energy, $E_{form}$. Under near-equilibrium growth condition, the probability of forming CdTe clusters can be estimated, within the canonical ensemble assumption, by

$$P_{cluster} = \frac{1}{ZN_{cluster}} e^{-E_{form}/kT}, \qquad (1)$$

where $Z$ is the canonical partition function, $N_{cluster}$ is the number of atoms in a cluster, and $kT$ is the thermal energy of 0.05 eV corresponding to a growth temperature of 300 °C.

To proceed, we consider the nucleation of stable clusters in the early stage of growth, as they are likely to determine the epitaxial relationship of the film with respect to the substrate [20]. These clusters should be smallest possible, yet large enough in size not to rotate by thermal excitation. Since our focus is on the growth on a substrate, here we consider only flat clusters that can best wet the substrate, whose atomic structures (in a top view) are shown in Fig. 1(a) for 1-ML- and 2-ML-thickness, respectively. We will pay special attention to the 1, 7, and 19 hexagon-ring clusters (termed 1R, 7R and 19R, respectively), as they are the most compact clusters in this size range with a least number of edge dangling bonds. In our definition, the A atoms of the binary AB compounds are in contact with the substrate. At 1-ML, the A/B atom ratios are 3:3, 12:12, and 27:27. At 2-ML, the ratios are 6:4, 24:19, and 54:46. Hence, stable 2-ML clusters are intrinsically A-atom rich. The atomic structures of bare NbSe$_2$ and graphene substrates are given in Fig. 1(b), showing that NbSe$_2$ has only a threefold rotational symmetry, whereas graphene has a sixfold rotational symmetry.

Epitaxial growth can happen either laterally or vertically (Fig. S1 of Supplemental Material). To determine which cluster(s) are vital to the experimentally observed epitaxial relationship, we have calculated numerous clusters. It is found that clusters with A = Cd have noticeably stronger interfacial binding energies than clusters with A = Te. Hence, herein we will discuss only clusters with Cd contacting the substrate. Figures 2(a-b) show the calculated formation energy for 1R, 7R, 19R clusters and bulk films ($\infty$R) with 1-, 2-, and 4-ML thicknesses, which reveals that the formation energies of CdTe-on-NbSe$_2$ are generally lower than those of CdTe-on-G. The corresponding atomic structures are described in Figs. S2-S3 of Supplemental Material. As the 1-ML clusters are stoichiometric, their formation energy is independent of the growth conditions. On the other hand, due to the dependence on the chemical potential, the formation energy of 2-ML clusters spans an energy range, as represented by the vertical bar in Figs. 2(a-b). Because all the 2-ML clusters are Cd-rich, therefore it makes sense to use the lower bounds in the following discussions. It shows that the 2-ML clusters are more stable than the 1-ML clusters, which also makes sense as the 2-ML clusters structural-wise are much closer to bulk films.

Figures 2(a-b) show that, regardless the choice of the substrate, the smallest 1R clusters have relatively high energies and hence prefer to grow laterally. As the size increases, however, results for CdTe-on-NbSe$_2$ can be quite different from those for CdTe-on-G. In the former case, the energy differences between the 7R and 19R clusters are rather small, e.g., $\Delta E_{form}$ < 0.02 eV/atom for both 1- and 2-ML clusters. The difference between the 2-ML 7R cluster and 2-ML film is also in the same energy range. Even when a thicker 4-ML film forms, the energy lowering from that of the 2-ML film is < 0.02 eV/atom. Using Eq. (1) and Fig. 2(a), we determine that the 2-ML and 1-ML 7R clusters are the most abundant clusters during the initial growth stage. Hence, they should be used to study the epitaxial relationship between CdTe and NbSe$_2$. In contrast, in the case of CdTe-on-G, $\Delta E_{form}$ is on the order of 0.1 eV/atom between 2-ML 7R and 19R, between 2-ML 19R and 2-ML film, and between 2-ML film and 4-ML film. From Eq. (1) and Fig. 2(b), the most abundant clusters on graphene substrate are 4-ML and 2-ML 19R clusters, followed by 2-ML and 1-ML 7R clusters.

To determine the epitaxial relationship of the cluster with respect to the substrate, we need to examine how the total energy of the system, $E_{tot}$, varies with in-plane orientation angle of the cluster. This is shown in Fig. 2(c) for 2-ML CdTe$_{7R}$-on-NbSe$_2$, where 0° is defined as $[1\bar{1}0]$CdTe

|| [2$\bar{1}\bar{1}$0]Substrate. It takes about 0.2 eV/cluster to rotate the cluster by 30° but the structure is highly unstable. It will take 0.29 eV/cluster to rotate the cluster by 60°, which is 0.13 eV/cluster higher in energy. Given that $kT \approx 0.05$ eV, the barrier here is large enough to fix the epitaxial relationship of the clusters before they grow into the larger and less-rotatable 19R clusters. Hence, we conclude that CdTe should take the parallel epitaxy (namely 0°). It appears that using 1-ML CdTe$_{7R}$-on-NbSe$_2$, the same conclusion can also be reached. In contrast, for 2-ML CdTe$_{7R}$-on-G in Fig. 2(d), it only takes 0.08 eV/cluster to rotate the cluster by 30°, which is 0.05 eV/cluster higher in energy. This barrier is unlikely to fix the epitaxial relationship for it is too close to the thermal energy of 0.05 eV. Hence, both 0° and 30° epitaxies exist for CdTe on graphene. As clusters grow larger, their rotation becomes less likely, as our calculation shows a much larger barrier of 0.39 eV/cluster when the size is at 19R. This is indeed what was observed by experimental XRD pole figure measurements [16, 17] showing a significantly larger FWHM for CdTe-on-G (~14°) than that for CdTe-on-NbSe$_2$ (~5°).

To understand the physical origin of the different energy landscape between NbSe$_2$ and graphene, next we consider the change in the electron density ($\Delta\rho$) due to the interface formation, shown in Fig. 3 for 2-ML CdTe$_{7R}$ cluster on (a-c) NbSe$_2$ and (d) graphene, respectively. It shows that, while $\Delta\rho$ associated with the interface formation is significant for CdTe-on-NbSe$_2$, the same can be neglected for CdTe-on-G. These results are consistent with our Bader analysis, which shows a total of 4.76-electron transfer from CdTe to NbSe$_2$ for CdTe$_{7R}$-on-NbSe$_2$, but only a total of 0.92-electron transfer for CdTe$_{7R}$-on-G. The smaller transfer to graphene is consistent with the fact that vdW force is rooted in dipole-dipole interaction that does not involve much transfer of electrons.

In the following, therefore, we will focus on 2-ML CdTe$_{7R}$-on-NbSe$_2$ for a possible mechanism of enhanced vdW epitaxy. It appears that one can understand the electron transfer qualitatively by examining the planar-averaged $\overline{\Delta\rho}(z)$ in Fig. 3(a) with three characteristic features: (1) a depletion of electrons near the Cd atoms at the bottom surface of the cluster, (2) an accumulation of electrons in the interfacial region between Cd and Se, and (3) a weakening of the Nb-Se bonding beneath the interface. One can find more details in the side view in Fig. 3(b) where the non-averaged $\Delta\rho(\vec{r})$ is shown. Electrons depleted from surface Cd atoms accumulate in the region between the Cd and Se atoms at the interface, as can be seen in Fig. 3(b). In accordance, electrons in the bonding states in the next layer between Nb and Se atoms retract somewhat back to the respective atoms. Figures 3(c-d) plot, in a top view, a positive (red) and a negative (blue) isosurface of $\Delta\rho(\vec{r})$ for CdTe-on-NbSe$_2$ and CdTe-on-G, respectively, using the same contour values. These plots show that, while electron rearrangement for CdTe-on-NbSe$_2$ is extensive, it can be neglected for CdTe-on-G, which of course is in line with our expectation.

To understand what happens in CdTe-on-NbSe$_2$, we plot the atomic site-decomposed density of state (DOS) in Figs. 4(a) for interfacial atoms, showing a noticeable orbital interactions between the Cd and Se states (cf. Fig. S4 of Supplemental Material for DOS over a wider energy range). As a result of the interactions, the originally half-occupied Cd-*sp* states are fully emptied and subsequently pushed up in energy, while the nearly-fully-occupied Se-$p_z$ states are fully occupied and subsequently pushed down. These occupation changes enable dative bond formation between empty cation state and doubly-occupied anion state. Figure 4(b) summarizes schematically the main results in Figs. 3 and 4(a) in terms of the above two steps: (I) an electron transfer across the interface and (II) a level repulsion through orbital interaction and the subsequent dative bond formation.

Strictly speaking, the two-step chemistry in Fig. 4(b) deviates from the standard dative-bond picture, as the bond formation here relies on the emptying of Cd dangling-bond states, which becomes possible only because of the presence of Nb non-bonding *d* states [see Fig. 4(a)]. These states serve as a reservoir to accept excess electrons. If NbSe$_2$ were a semiconductor, for example, the Se lone pair would be fully occupied. The 2-ML CdTe$_{7R}$ cluster, on the other hand, has an excess of $(24 - 19) \times 2 = 10$ electrons (distributed over 12 Cd atoms with 18 dangling bonds). Because the fully-occupied Se lone pairs have no room to accept these electrons, the strength of the dative bonds will be greatly reduced. In reality, the Nb *d* states not only accept electrons from the CdTe states but also from the Se-$p_z$ states. Therefore, *on the appearance*, electron transfer takes place between Cd dangling-bond and Se-$p_z$ lone-pair states, as schematically illustrated in Fig. 4(b) [21].

For CdTe-on-G, in principle there can also be an electron transfer from the CdTe to metallic graphene states, which would also suggest a significantly enhanced vdW binding through dative bond formation which, however, does not happen. This is because NbSe$_2$ is considerably more electronegative than graphene, as evidence by their different workfunctions: $\Phi = 5.9$ and 4.6 eV for NbSe$_2$ and graphene [22, 23], respectively. These values may be contrasted to the electron affinity ($\chi = 4.3$ eV) for CdTe [24]. Hence, the driving force for electron transfer from CdTe to NbSe$_2$ is much greater than that to graphene, as having also being revealed by our Bader analysis.

Note that, while an electron transfer can enhance interfacial binding, it may not enhance vdW epitaxy, as the latter requires an enhancement of directional bonding. It is the second step, namely, the formation of dative bonds that plays the sole role in enhanced vdW epitaxy. On this note, graphene is not a good substrate for significantly enhanced vdW epitaxy, also because its $p_z$ orbital is unsuited for dative bond formation due to the formation of stable aromatic bonding network. The dative bond energy for CdTe-on-NbSe$_2$ (1.3 eV/CdSe) is less than the standard Cd-Se bond strength of 2.7 eV/CdSe, deduced from the cohesive energy of wurtzite CdSe [25]. One may covert this value into an interfacial binding energy of 1.88 eV/CdTe at the CdTe-NbSe$_2$ interface, which is, however, five times as large as that of 0.38 eV/CdTe at the CdTe-G interface.

The mechanism for enhanced vdW binding discussed above is clearly beyond just CdTe-NbSe$_2$. It should apply broadly to other substrates as well, as long as the interfacial dative bond formation can be activated by a removal of excess electrons at the cation sites. For example, it was shown that CdTe growth on semiconducting MoTe$_2$ and WSe$_2$ [26] has poor in-plane orientation. We can remove the excess electrons on interfacial Cd atoms, whereby activating the dative bond formation, by introducing holes through *p*-type doping.

In summary, an electronic theory that differentiates the epitaxy of 3D semiconductors on seemingly identical vdW layered substrates is proposed. Application to the prototypical CdTe epitaxy on NbSe$_2$ and graphene, by means of first-principles calculation, reveals the origin for their differences, in particular, the enhanced vdW epitaxy of CdTe on NbSe$_2$ can now be understood as the formation of interfacial dative bonds. Unlike a standard dative-bond mechanism, however, here electron transfer between surface Cd dangling-bond states and non-bonding NbSe$_2$ states must precede the bond formation. Understanding and harnessing this unexplored regime of epitaxy is at the cusp of materials physics, which offers great potentials not only to grow epitaxial 3D crystals and superlattices on 2D vdW substrates, but also to understand how the bonding chemistry operates in naturally occurring layered crystals, which can be vital for novel crystal design.

We thank Damien West for helpful discussions. This work was supported by the National Science Foundation Award No. 1305293. The supercomputer time sponsored by NERSC under DOE contract No. DE-AC02-05CH11231 and the CCI at RPI are also acknowledged.

**Figures**

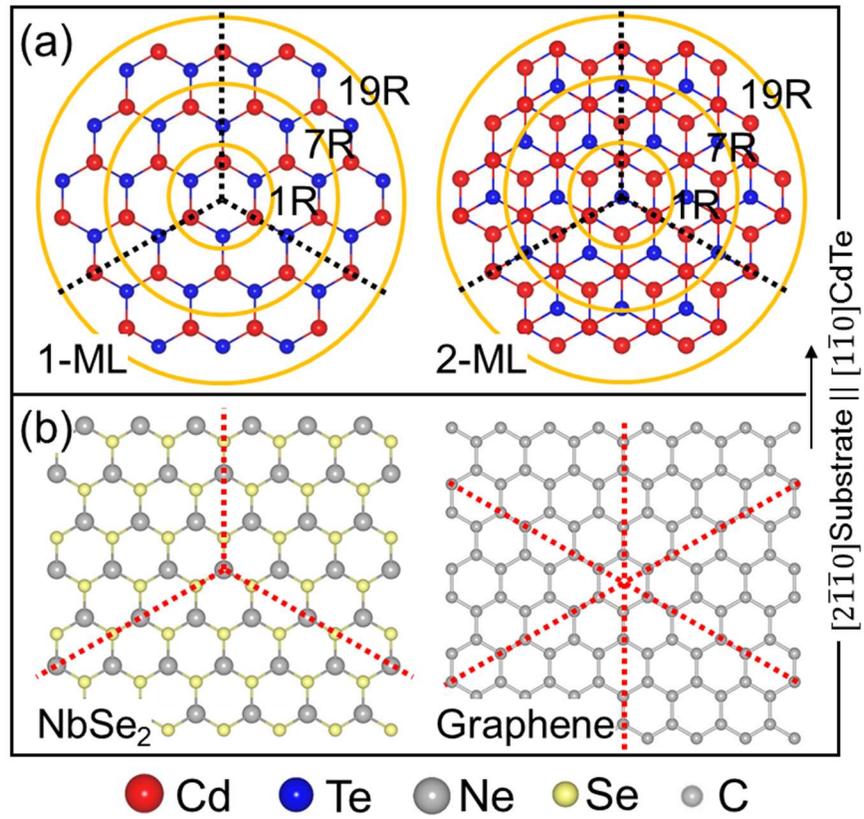

**FIG. 1.** Atomic structures of (a) 1-ML and 2-ML CdTe(111) clusters, in which the 1, 7, and 19 hexagon rings have been marked. (b) NbSe₂ and graphene substrates. Dotted lines show the various crystal symmetries.

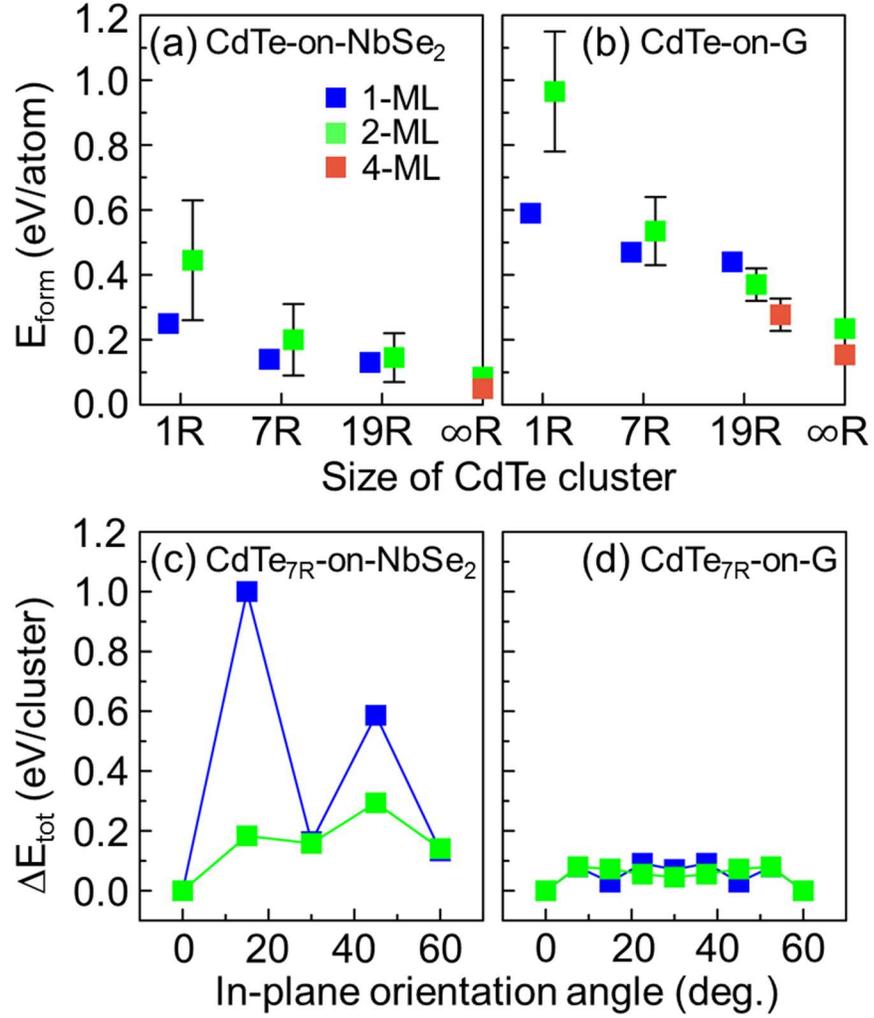

**FIG. 2.** (a-b) Formation energies for stable CdTe clusters at different thicknesses on (a) NbSe$_2$ and (b) graphene (G) substrates. Energy for the 2-ML-thick clusters is generally a function of the growth condition with lower bound = Cd-rich and upper bound = Cd-poor. (c-d) Total energy difference as a function of the in-plane orientation angle between the CdTe$_{7R}$ cluster and (c) NbSe$_2$ and (d) graphene substrate, respectively.

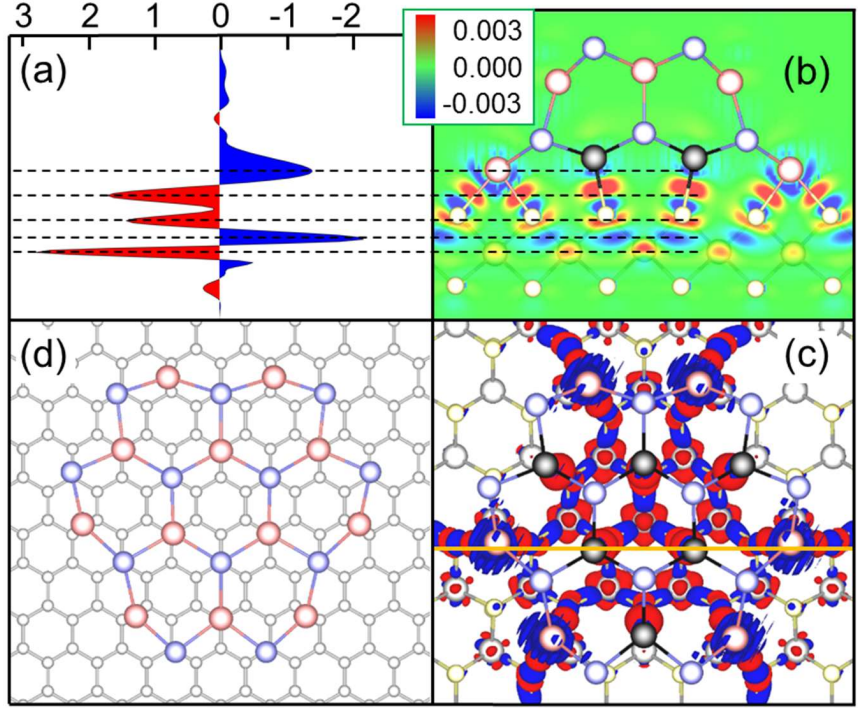

**FIG. 3.** Electron density change ($\Delta\rho$) near the interface between 2-ML CdTe$_{7R}$ and (a-c) NbSe$_2$ and (d) graphene substrate. (a) Planer-averaged $\overline{\Delta\rho}(z)$ (e/Å). (b) Side view and (c) top view of $\Delta\rho(\vec{r})$ (e/Å$^3$): red = accumulation and blue = depletion. For clarity, in (b) only atoms near the selected plane, denoted by the orange line in (c), are shown. In (c-d), on the other hand, only the bottom layer of the CdTe cluster is shown, with contour value of ±0.003 e/Å$^3$. To make a distinction between the corner and central atoms, the 6 central Cd atoms are darkened.

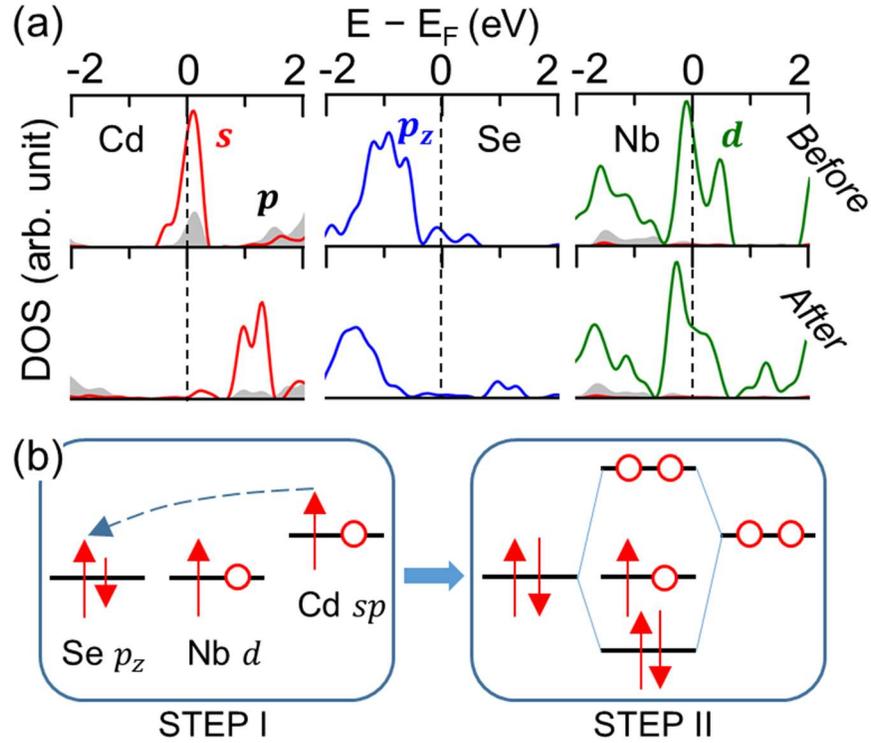

**FIG. 4.** (a) Site-decomposed DOS for interfacial Cd, Se, and Nb atoms *before* and *after* the formation of the interface between 2-ML CdTe$_{7R}$ cluster and NbSe$_2$ substrate. (b) A schematic diagram summarizing the electron transfer in step I and dative bond formation through level repulsion between interfacial Cd and Se states in step II. A shorter spin arrow in (b) indicates that in that spin channel, there is still room to accept electrons.

# Supplemental Material for
# Enhanced van der Waals epitaxy via electron transfer-enabled interfacial dative bond formation

Weiyu Xie[1], Toh-Ming Lu[1], Gwo-Ching Wang[1], Ishwara Bhat[2], and Shengbai Zhang[1]*

[1]*Department of Physics, Applied Physics & Astronomy,*

*Rensselaer Polytechnic Institute, Troy, NY 12180, USA*

[2]*Department of Electrical, Computer and Systems Engineering,*

*Rensselaer Polytechnic Institute, Troy, NY 12180, USA*

*Corresponding author, Shengbai Zhang. Phone: (518) 276-6127; Email: zhangs9@rpi.edu

**Method**

First-principles density functional theory (DFT) calculations are carried out within the generalized gradient approximation (GGA) using the Perdew-Burke-Ernzerhof (PBE) functional [1] and the projector augmented wave (PAW) method [2], as implemented in the Vienna *ab initio* Simulation Package (VASP) [3]. In order to describe the vdW interactions in the system, Grimme's DFT-D3 method [4, 5] is used. We employ a plane-wave basis set with a kinetic energy cutoff of 400 eV, and a force convergence criterion of 0.05 eV/Å for structural optimization. Most of the studies involve cluster-on-substrate systems in a periodic slab with a 15-Å vacuum region between slabs and an 11-Å lateral separation between the edges of adjacent clusters. For cluster size < 24 atoms and NbSe$_2$ substrate size < (8×8), the Brillouin zone integration is performed using a 2×2×1 $k$-point mesh. Since graphene lattice is noticeably smaller than NbSe$_2$, the same $k$-point mesh is used for graphene substrate size < (11×11). For larger systems, only Γ-point is used. Bader analysis based on a grid-based algorithm [6-8] is carried out using the DFT charge density.

Formation energy of the interface is calculated by

$$E_{\text{form}} = E_{\text{tot}} - \sum n_i \mu_i, \quad (S1)$$

where $E_{\text{tot}}$ is the total energy of the interface, $n_i$ is the number of atoms of species $i$ in the supercell, and $\mu_i$ is the chemical potential of the atom $i$ with respect to that of the elemental phase ($\mu_i^{\text{el}}$) by $\mu_i = \mu_i^{\text{el}} + \Delta\mu_i$. The chemical potentials of $\mu_{\text{Cd}}^{\text{el}}$ and $\mu_{\text{Te}}^{\text{el}}$ are given by bulk Cd (hexagonal, space group $P6_3/mmc$) [9] and Te (trigonal, space group $P3_121$) [10], respectively. Depending on the experimental growth conditions, $\mu_i$ can vary while maintaining the thermodynamic relation:

$$\Delta\mu_{\text{Cd}} + \Delta\mu_{\text{Te}} = \Delta H_f(\text{CdTe}) = -0.70 \text{ eV}, \quad (S2)$$

where $\Delta H_f(\text{CdTe})$ is the formation energy of CdTe (cubic, space group $P\bar{4}3m$) [9].

**Figures**

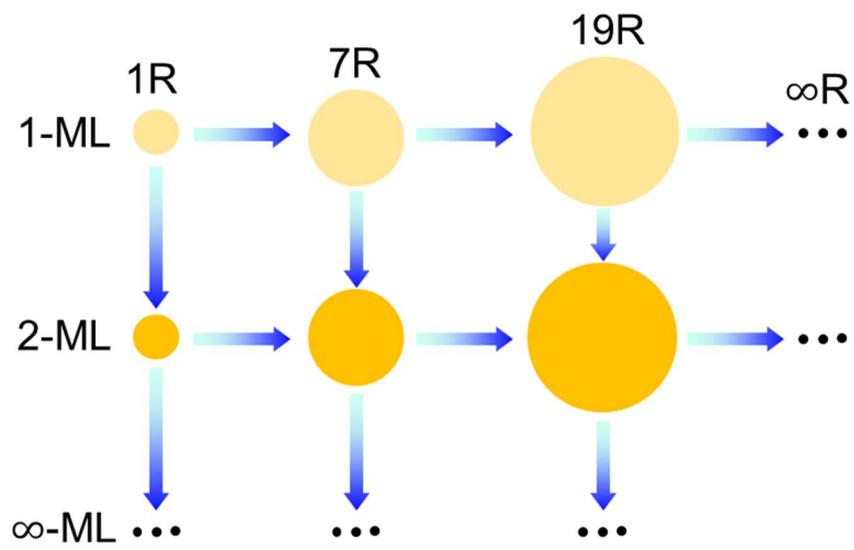

**FIG. S1.** Schematic illustration for the growth of CdTe clusters on a substrate: from left to right, the size of the cluster increases; from top to bottom, the thickness of the cluster increases.

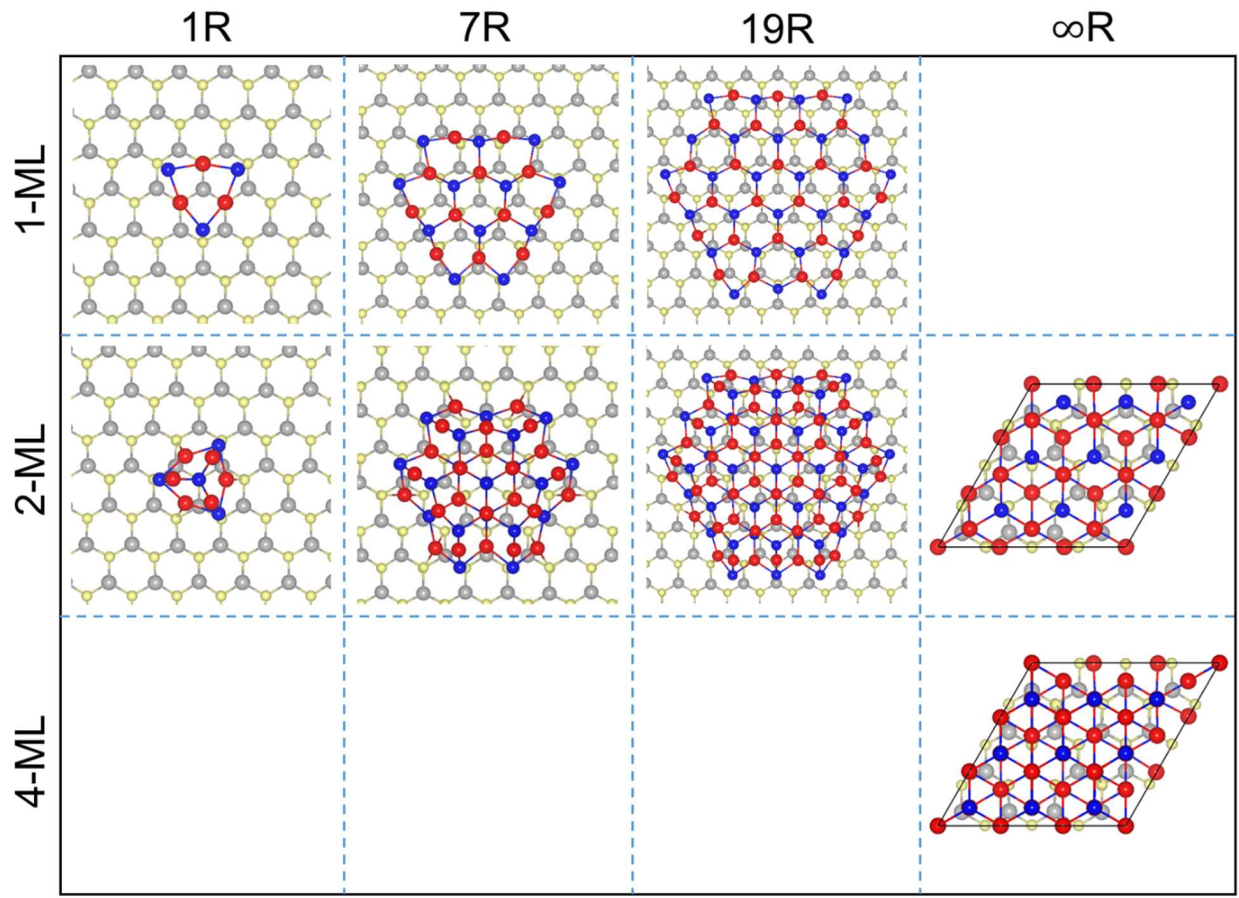

**FIG. S2.** Fully-relaxed stable CdTe clusters of variable thicknesses on NbSe$_2$ substrate, corresponding to Fig. 2(a). Both in-plane rotation and in-plane translation of CdTe clusters relative to the substrate are considered for structural optimization.

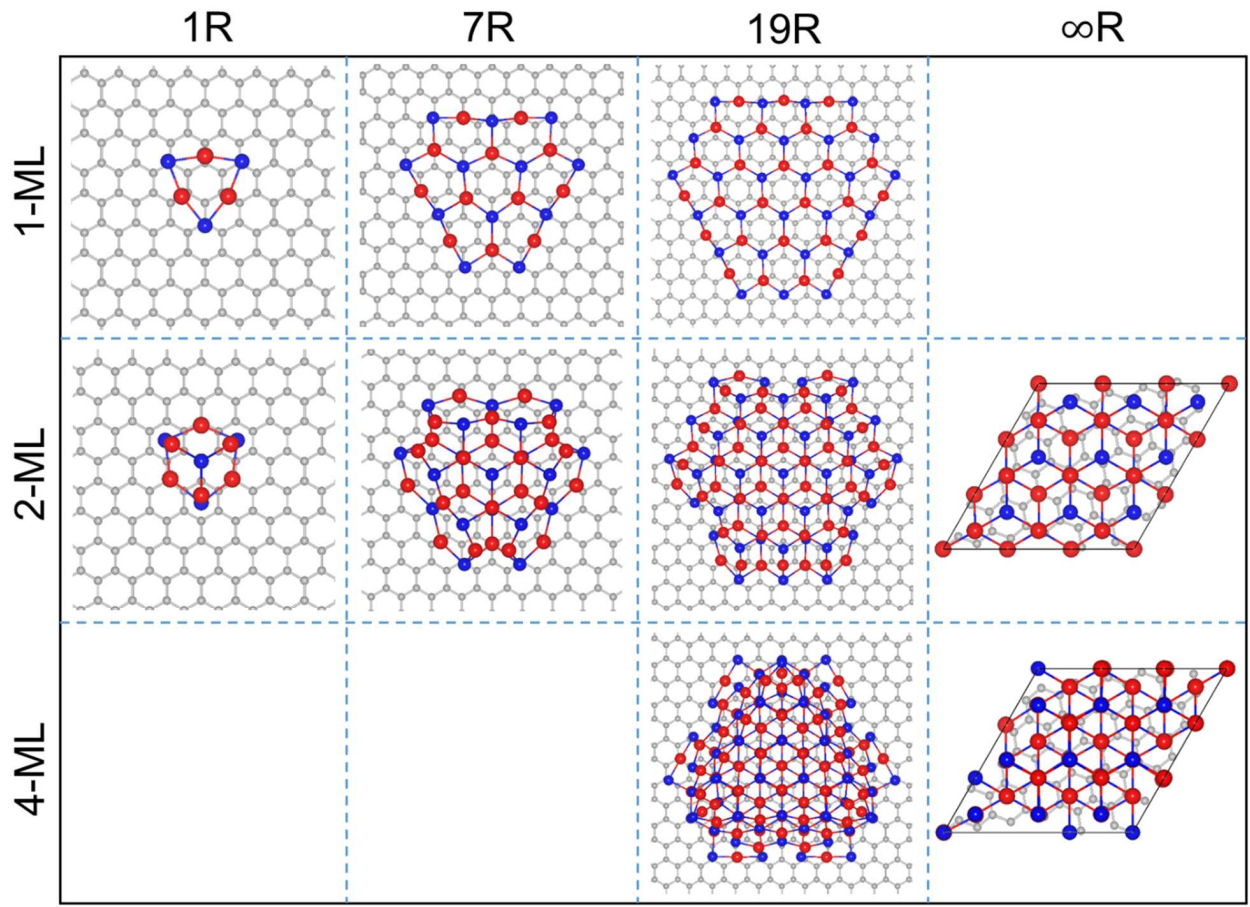

**FIG. S3.** Fully-relaxed stable CdTe clusters of variable thicknesses on graphene substrate, corresponding to Fig. 2(b). Both in-plane rotation and in-plane translation of CdTe clusters relative to the substrate are considered for structural optimization.

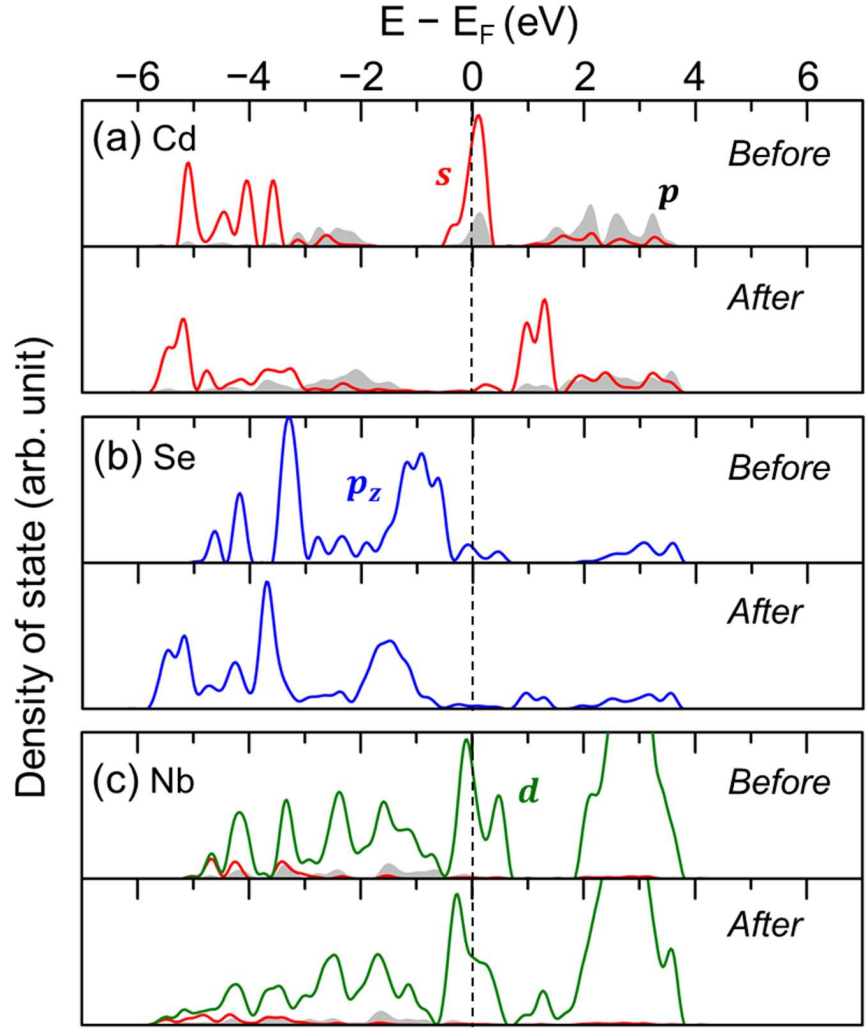

**FIG. S4.** Site-decomposed DOS in Fig. 4 over a wider energy range. Red curve is for Cd-$s$ state, shaded grey is for Cd-$p$ states, blue curve is for Se-$p_z$ state, and green curve is for Nb-$d$ states.